\documentclass[aps,twocolumn,pra,tightenlines,floatfix,showpacs,superscriptaddress,longbibliography]{revtex4-1}
\usepackage{graphicx}
\usepackage{amsmath}
\usepackage{amssymb}
\usepackage{times}
\usepackage{epstopdf}
\usepackage[english]{babel}
\usepackage{natbib}
\usepackage{hyperref}
\usepackage{physics}
\hypersetup{colorlinks=true, citecolor=blue, linkcolor=blue}
\usepackage{color}
\usepackage{tablefootnote}
\usepackage{braket}

\newcommand{\dg}{\ensuremath{^\dagger}}

\newcommand{\nn}{\nonumber}

\begin{document}
\title{Quantum Markov Chain Monte Carlo with \\ Digital Dissipative Dynamics on Quantum Computers}
\author{Mekena Metcalf}
\email{mmetcalf@lbl.gov}
\affiliation{Computational Research Division, Lawrence Berkeley National Laboratory, Berkeley, CA 94720, USA}
\author{Emma Stone}
\affiliation{Department of Physics, North Carolina State University, Raleigh, North Carolina 27695, USA}
\author{Katherine Klymko}
\affiliation{Computational Research Division, Lawrence Berkeley National Laboratory, Berkeley, CA 94720, USA}
\author{Alexander F. Kemper}
\affiliation{Department of Physics, North Carolina State University, Raleigh, North Carolina 27695, USA}
\author{Mohan Sarovar}
\affiliation{Extreme-scale Data Science and Analytics, Sandia National Laboratories, Livermore, CA 94550, USA}
\author{Wibe A. de Jong}
\email{wadejong@lbl.gov}
\affiliation{Computational Research Division, Lawrence Berkeley National Laboratory, Berkeley, CA 94720, USA}

\begin{abstract}
Modeling the dynamics of a quantum system connected to the environment is critical for advancing our understanding of complex quantum processes, as most quantum processes in nature are affected by an environment.
Modeling a macroscopic environment on a quantum simulator may be achieved by coupling independent ancilla qubits that facilitate energy exchange in an appropriate manner with the system and mimic an environment.
This approach requires a large, and possibly exponential number of ancillary degrees of freedom which is impractical.
In contrast, we develop a digital quantum algorithm that simulates interaction with an environment using a small number of ancilla qubits. By combining periodic modulation of the ancilla energies, or spectral combing, with periodic reset operations, we are able to mimic interaction with a large environment and generate thermal states of interacting many-body systems.
We evaluate the algorithm by simulating preparation of thermal states of the transverse Ising model. 
Our algorithm can also be viewed as a quantum Markov chain Monte Carlo (QMCMC) process that allows sampling of the Gibbs distribution of a multivariate model. To demonstrate this we evaluate the accuracy of sampling Gibbs distributions of simple probabilistic graphical models using the algorithm. 
\end{abstract}
\maketitle
\section{Introduction}
Much of simulation science is built upon modeling a system interacting with an environment in order to capture complex dissipative and relaxation dynamics. In addition, the use of engineered environmental interactions to prepare equilibrium states of interacting systems has a long history in computing. 
However, environments tend to be very large, and directly simulating a system coupled to an environment using classical or quantum simulation requires approximations and many extra degrees of freedom to serve as 
a source of entropy.
Sampling from difficult distributions like a multivariate Gibbs distribution on quantum computers has been proposed using Metropolis sampling algorithms that reduce time complexity \cite{temme_quantum_2011,yung2012quantum,szegedy2004quantum, wocjan2008speedup,wild2020quantum}, variational algorithms that are possibly well-suited to near-term quantum computers but have a classical optimization overhead \cite{Verdon_VQT,endo2020variational}, thermal-field double states \cite{Zhu_TFD_2020,Francis2020}, and quantum imaginary time-evolution to implement the minimally entangled typical thermal state (METTS)~\cite{white2009minimally,stoudenmire2010minimally} sampling algorithm on quantum computers \cite{Motta_2020, Sun_QITE_2020}.  Our approach is different, as we engineer an open-quantum system with the desired thermal/Gibbs state as the fixed point of evolution.  

Thermal states can be prepared by modeling open-quantum systems if the system and environment are weakly coupled, if the dynamics are ergodic, and if the energy exchange is detailed balanced \cite{breuer_theory_2002, shabani_artificial_2016, metcalf2020engineered}. 
However, it is not straightforward to guarantee that these conditions are met by an environment that is engineered from ancilla degrees of freedom, and moreover, the number of required ancilla degrees of freedom could scale exponentially with the degrees of freedom in the system to be thermalized (since the number of energy eigenstates scales exponentially with degrees of freedom).   
We address these difficulties 
by adding time-dependence to the ancilla qubits that approximate the environment, and modulating the energy of these ancilla qubits across the system energy spectrum in a suitable manner -- a process termed \emph{spectral combing}. This procedure allows us to engineer the necessary conditions for thermalization with a number of ancilla that does not need to scale exponentially with the system size. The algorithm we develop in this paper can be viewed as a digital version of the analog thermalization algorithm developed in Ref. \cite{metcalf2020engineered}, which relied on the same principles. Our algorithm is a method to emulate quantum Markov chain Monte Carlo (QMCMC) methods~\cite{levin2017markov} on quantum devices \cite{montanaro2015quantum,yung2012quantum}, and thus provides a general route to sample from complex probability distributions corresponding to the stationary states of Markov chains. In the case that the eigenvalues of the Hamiltonian are non-degenerate and can be identified by a label, the algorithm reduces to the classical Markov chain mapped to a quantum Hamiltonian~\cite{temme_quantum_2011}.  
An important application will be the preparation of thermal, initial states to simulate chemical and molecular reaction pathways~\cite{von2020quantum} in large systems (where the classical dynamics become prohibitively expensive) at arbitrary temperatures. 
Our algorithm is designed to obtain thermal distributions of statistical physics problems with great accuracy. In the case of degenerate quantum algorithms we find our algorithm is able to obtain thermal distributions. The thermal state of Hamiltonians fully degenerate levels has greater error due to the violation of detailed balance. For concreteness, we focus on spin systems, however, the general approach can be used to thermalize any many-body system whose Hamiltonian can be encoded in a many-qubit Hamiltonian.

Spectral combing using auxiliary qubits that exchange energy with the principal qubits periodically in time has been proposed for digital and analog unitary evolution to obtain ground states and thermal states \cite{kaplan_ground_2017, polla2019quantum}. Our approach in Ref~\cite{metcalf2020engineered} used spectral combing to engineer detailed balance conditions on analog quantum computers. By engineering the detailed balance conditions we can obtain high-accuracy, approximate thermal distributions on quantum devices~\cite{shabani_artificial_2016}. We extend our previous method to digital quantum computers, and devise an algorithm that mimics the interaction of a system with a finite temperature macroscopic bath using time-dependent ancilla qubits. Our algorithm is founded on the three criteria needed for the thermal state to be the unique fixed point of evolution: Born-Markov approximation, ergodic dynamics, and detailed balance transitions \cite{breuer_theory_2002}. We test the performance of our algorithm by numerically evaluating the distance between the steady-state solution and genuine, thermal distribution for a transverse field Ising model (TFIM) and by computing the error in computing the thermal magnetization in the transverse direction. We also evaluate the performance of the algorithm on the problem of sampling multivariate Gibbs distributions over binary random variables with constraints generated by Erd\'os R\'enyi random graphs.

\section{Theoretical Foundation}

\begin{figure*}
  \includegraphics[width=\textwidth]{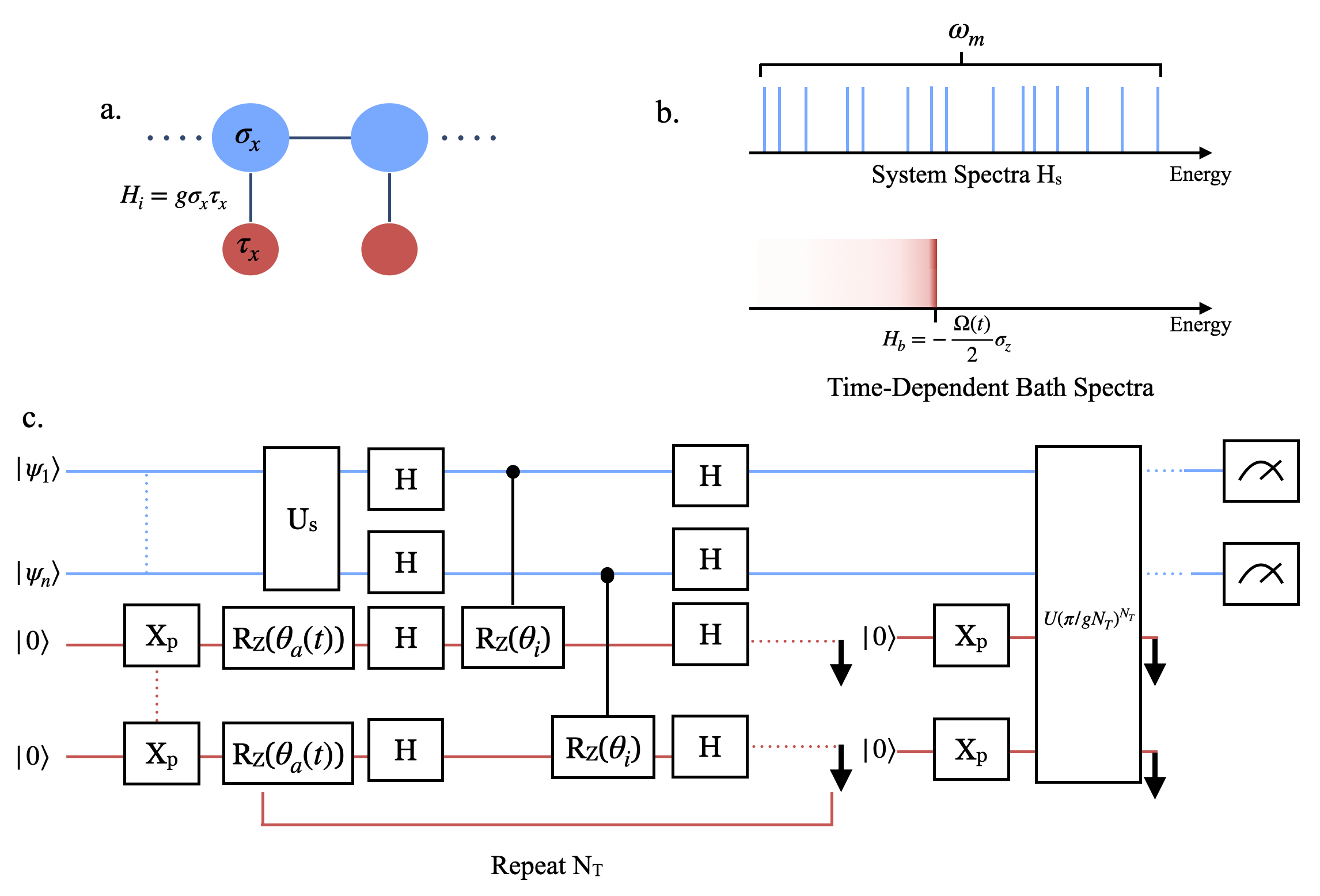}
  \caption{a) Principal qubits (blue) locally connected to ancilla qubits (red). b) Time-dependent ancilla frequency combs the system energy spectra and resonantly exchanges energy with different energy transitions in the system at different times c) Quantum circuit to implement the interaction cycle dynamical map in Eq. \ref{eq:map} for arbitrary system size. $U_s = e^{-iH_s \pi/g N_T}$ and $U(\pi/gN_T)$ is what we denote $W_t$ in the main text evaluated at some time $t$. The phase rotation with angle $\theta_a(t) = -\Omega(t)\pi /g N_T$ associated with the ancilla Hamiltonian does not affect the dynamics until $U(\pi /gN_T)^2$.  The system-bath interaction yields an angle $\theta_i = 2\pi/N_T$. $X_p$ is a probabilistic $X$ gate, which is applied with probability $(1-p_0(t))$, see text for details.}
  \label{Fig:AlgIll}
\end{figure*}

Modeling finite temperature physics of quantum systems requires defining a composite Hilbert space of the system and a bath $\mathcal{H} = \mathcal{H}_s\otimes \mathcal{H}_b$. The total Hamiltonian defines the system $H_s$, the bath $H_b$, and the interaction between them $H_i$,
\begin{equation}
\label{Eq:Ham}
    H = H_s\otimes I + I \otimes H_b + H_i.
\end{equation}
We can now express the time-evolution operator of the total Hamiltonian 
\begin{equation}
    \rho(t) = U(t)\rho(0) U^\dagger(t),
\end{equation}
\begin{equation}
    U(t) = e^{-iHt}, 
\end{equation}
where we set $\hbar = 1$. 

If the system and bath are weakly coupled, $\vert\vert H_i \vert\vert \ll \vert\vert H_b \vert\vert \ll, \vert\vert H_s \vert\vert \ll$, and the bath is fast equilibriating, the influence of the system on the bath state is negligible. We can make the Born approximation, defining the state of the system-bath at $t=0$ as an uncorrelated product state $\rho(0) = \rho_s(0)\otimes \rho_b$. Tracing out the bath degrees of freedom the state of the system at time $t$, 
\begin{equation}
    \rho_s(t) = \text{Tr}_b \left(U(t,0)\left[\rho_s(0)\otimes \rho_b\right] U^\dagger(t,0)\right), 
\end{equation} defines a completely positive, trace preserving (CPTP) dynamical map. This dynamical map is what we aim to implement on a quantum computer. Under the conditions that the system dynamics is ergodic and that the bath-induced transitions between system eigenstates are detailed balanced \cite{breuer_theory_2002}, the unique fixed point of evolution is the thermal state
\begin{equation}
    \rho_s(t) = \rho_{th} = \frac{e^{-\beta H}}{\text{Tr}\left(e^{-\beta H}\right)},
\end{equation}
where $\beta=1/k_BT$, and $T$ is the equilibrium temperature.
Using insights from Ref. \cite{metcalf2020engineered} we demonstrate how these conditions can be satisfied by engineering a bath consisting of locally coupled, time-independent ancilla spins.
\section{Quantum Algorithm}
 The algorithm weakly couples a collection of periodically modulated ancilla qubits (which function as the ``bath'') to the qubits in an arbitrary multi-qubit system Hamiltonian ($H_s$) as such,
\begin{align}
    H_b(t) &= - \sum_{m=1}^M \frac{\Omega(t)}{2}\tau^m_z \\
    H_i &= \sum_{m=1}^M g \left(\sigma_x^{\lambda(m)}\otimes \tau_x^m\right)\\
    H(t) & = \left(H_s\otimes I_b\right) + \left(I_s\otimes H_b(t)\right) + H_i,
\end{align}
where $\tau_\alpha^m$ $(\alpha = x,y,z)$ are the Pauli spin operators acting on ancilla $m$, and $\sigma_\alpha^{\lambda(m)}$ are spin operators for the principal spin coupled to ancilla $m$, see Fig.~\ref{Fig:AlgIll}(a). $\lambda(m)$ simply returns the index of the principal qubit that is coupled to ancilla qubit $m$. The function $\Omega(t) = \omega_m f(t) $ represents the time-dependent modulation of the bath qubits, and sweeps  across the system spectrum using a periodic function $f(t)$ with period $T_{cycle}$ in conjunction with an estimate of the spectral width of the system Hamiltonian, $\omega_m \approx \abs{E_{max}-E_{min}}$. Periodically modulating the energy of the {\it independent} ancilla spins enables an exchange of energy with different system frequency transitions at different times, see Fig.~\ref{Fig:AlgIll}(b). In the above Hamiltonian we have modeled the system-ancilla interaction through a $\sigma_x^{\lambda(m)}\otimes\tau_x^m$ interaction. This is not uniquely specified. This could be any other interaction that promotes energy exchange between the system and ancilla degrees of freedom, e.g., $\sigma_y^{\lambda(m)}\otimes\tau_y^m$. The important element is that the system portion of this interaction (e.g., $\sigma_x^{\lambda(m)}$) cannot commute with $H_s$. Finally, while not necessary, in the above model we have assumed that the system-ancilla coupling, $g$, is the same for all $m$ for simplicity.

Our approach in the following will be to approximate the above continuous evolution via a discretized Trotter evolution and develop a gate-based implementation of the time evolution However, crucial to the thermalization behavior is engineering the ancilla systems to mimic a macroscopic bath that is on average in a thermal state. To do this, we need to ensure that the ancilla qubits, whose local eigenbasis is the computational basis, are maintained in local thermal states over coarse timescales. We employ a (non-unitary) reset mechanism and a probabilistic application of a rotation in order to achieve this. The reset operation on all $M$ ancilla qubits is defined as $\mathcal{R}\rho \equiv \sum_{i=0}^{2^{M}-1} \dyad{0}{i}\rho \dyad{i}{0}$, where $\ket{i}$ is the $M$ qubit state encoding the binary representation of $i$. In more detail, the state of the ancilla spin that we wish to maintain at a time $t$, is the (time-dependent) thermal state 
\begin{align}
    \rho_{b}^{th}(t) &= p_0(t)\dyad{0}{0} + (1-p_0(t))\dyad{1}{1},\\
    p_0(t) &= \frac{e^{\beta \Omega(t)/2}}{e^{\beta \Omega(t)/2}+e^{-\beta \Omega(t)/2}}.
\end{align}
We prepare this state by reseting the ancilla spin to $\ket{0}$ and applying a $\tau_x^m$ rotation with probability $(1-p_0(t))$.


 
As shown in Ref. \cite{metcalf2020engineered} thermalization of the system relies on a separation of time-scales between the system-bath interaction and the bath relaxation -- essentially, the system should ``see'' all ancilla degrees of freedom in thermal states at the natural timescale of the system-bath interaction. Therefore, we distribute the ancilla reset and randomized preparation steps in such a way that there is a reset after a full cycle of system-ancilla interaction -- i.e., after an evolution time $T_g = \pi/g$. From another perspective, weak coupling of the system to the ancilla and the periodic resets of the ancilla qubits ensure there is no correlation between the ancilla spins, and phase differences will not lead to interference over the course of evolution. In order for effective thermalization of the system we require that the energy modulation of the ancilla be slower than the interaction timescale. More precisely, we require the following hierarchy of parameters:
\begin{align}
	\vert \frac{df(t)}{dt}\vert \ll g \ll \vert\vert H_{s}\vert\vert
	\label{eq:rate_hierarchy}
\end{align}

Putting this together, the operation on the composite system for a period of system-ancilla interaction is,
\begin{align}
	\rho(t+ T_g) = \left[ \mathcal{W}_t \circ \mathcal{F}_t\circ\mathcal{R} \right]\rho(t) ,
	\label{eq:map}
\end{align}
 where
 \begin{widetext}
\begin{align}
	\mathcal{W}_t\rho &= W_t \rho W_t\dg, \quad \textrm{with} \quad
	W_t = \left[ \left(\prod_{m=1}^M e^{-i g \left(\sigma_x^{\lambda(m)}\otimes \tau_x^m\right) \frac{T_g}{N_T}} \right) \left(e^{-iH_s \frac{T_g}{N_T}}\right) \left(\prod_{m=1}^M \left(e^{i\frac{\Omega(t)}{2}\tau_z^m \frac{T_g}{N_T}}\right)\right)\right]^{N_{T}}, \textrm{ and} \\
    \mathcal{F}_t\rho &= \otimes_{m=1}^M \left(p_0(t) \rho + (1-p_0(t)) \tau_x^m \rho \tau_x^m \right).
\end{align}
\end{widetext}
Note that we have suppressed the identity operations on subsystems being acted on trivially.
The operation $\mathcal{F}_t$ implements the probabilistic bit flip of the ancilla qubits.
The order of the unitary operators in $W_t$ describes the physical process of system-bath evolution, we first evolve by the system and bath Hamiltonians for a time $T_g/N_T$, followed by an application of the interaction. This is repeated for $N_{T}$ iterations, where $N_{T}$ is parameter to be chosen. Eq. \ref{eq:map} corresponds to a first-order Suzuki-Trotter discretization of the Hamiltonian dynamics described above coupled with the non-unitary operations of reset and probabilistic excitation of the ancilla qubits. The choice of $N_T$ dictates the error incurred in discretizing the continuous coherent evolution by Hamiltonian in Eq. \ref{Eq:Ham}. To achieve an error of $\epsilon$, we require $N_T = \mathcal{O}((3T_g\Lambda)^2/\epsilon)$ \cite{Childs_Ostrander_Su_2019}, where $\Lambda = \max(\Vert H_i\Vert, \Vert H_s\Vert, \Vert H_b\Vert)$. We note that one could implement the unitary portion of the dynamics, $W_t$, with higher order Suzuki-Trotter product formulas. This would increase the complexity of the gate sequence implementing the evolution but would allow for a smaller number of Trotter steps, $N_T$, while holding simulation error constant \cite{childs_theory_2019}. 

Eq. \ref{eq:map} provides a prescription for executing our thermalization algorithm on a quantum computer. A circuit representation of the sequence of operations is given in Figure 1(c). It is important to note that $\Omega(t)$, or more accurately $f(t)$, is assumed to be constant over an evolution time of $T_g$, consistent with the parameter hierarchy presented in Eq. \ref{eq:rate_hierarchy}. We introduce an additional parameter corresponding to the discretization of $f(t)$ -- we assume one period of $f(t)$, $T_{\rm cycle}$ is divided into $N_{\rm cycle}$ steps of size $T_g$; i.e., $T_{\rm cycle} = T_g \times N_{\rm cycle}$ sets the timescale of the sweep of the ancilla energies. The overall dynamics over one $T_{\rm cycle}$ is then:
\begin{align}
	\rho(t+T_{\rm cycle}) = \left[\prod_{k=0}^{N_{\rm cycle}}  \mathcal{W}_{t_k} \circ \mathcal{F}_{t_k} \circ \mathcal{R} \right] \left(\rho_s(t)\otimes \ket{0}_b\bra{0}\right),
	\label{eq:full_cycle}
\end{align}
with $t_k = kT_g$ and $\ket{0}_b\bra{0}$ denotes the state where all $M$ ancilla qubits are in their ground state. $N_{\rm cycle}$ is another parameter of choice, and we find that the appropriate choice depends on the details of the system being driven to thermal equilibrium. 

\section{Numerical Examination of QMCMC}
\begin{figure}
  \includegraphics[width=3.25in]{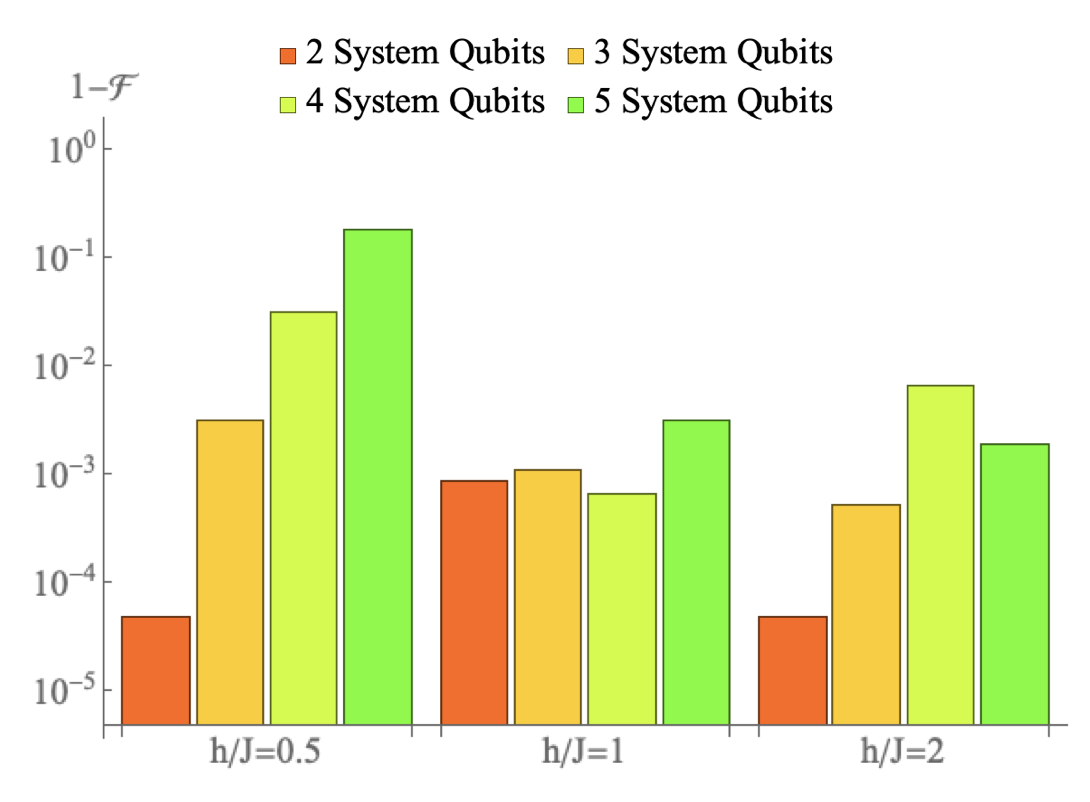}
  \caption{Infidelity of the QMCMC ground state with the genuine TFIM ground state with $h/J = 0.5$, $h/J = 1$, and $h/J = 2$. The TFIM has $N_s = 2-5$ principal qubits each independently coupled to a single ancilla spin prepared using $p_0 = 1$.}
  \label{Fig:TFIM_infid}
\end{figure}
\begin{figure}[ht]
  \includegraphics[width=0.4\textwidth]{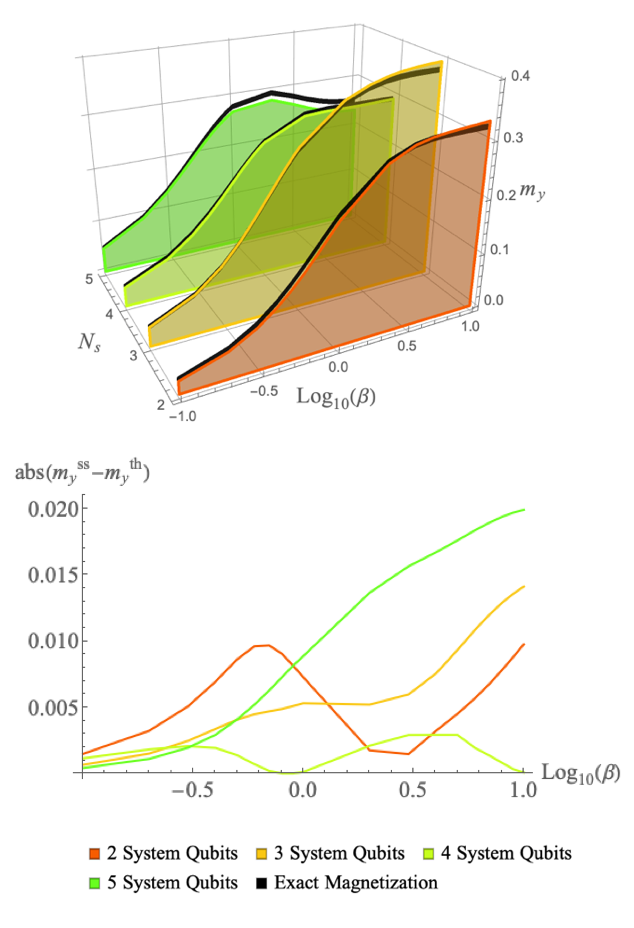}
  \caption{(upper) Transverse magnetization of the TFIM with $h/J=1$ using $N_s = 2-5$ principal qubits, as a function of the target (inverse) temperature compared to the exact magnetization shown by the black curves.(lower) Error between the exact magnetization and the magnetization of the algorithm steady-state.}
  \label{Fig:TFIMMag}
\end{figure}

\begin{figure*}
  \includegraphics[width=\textwidth]{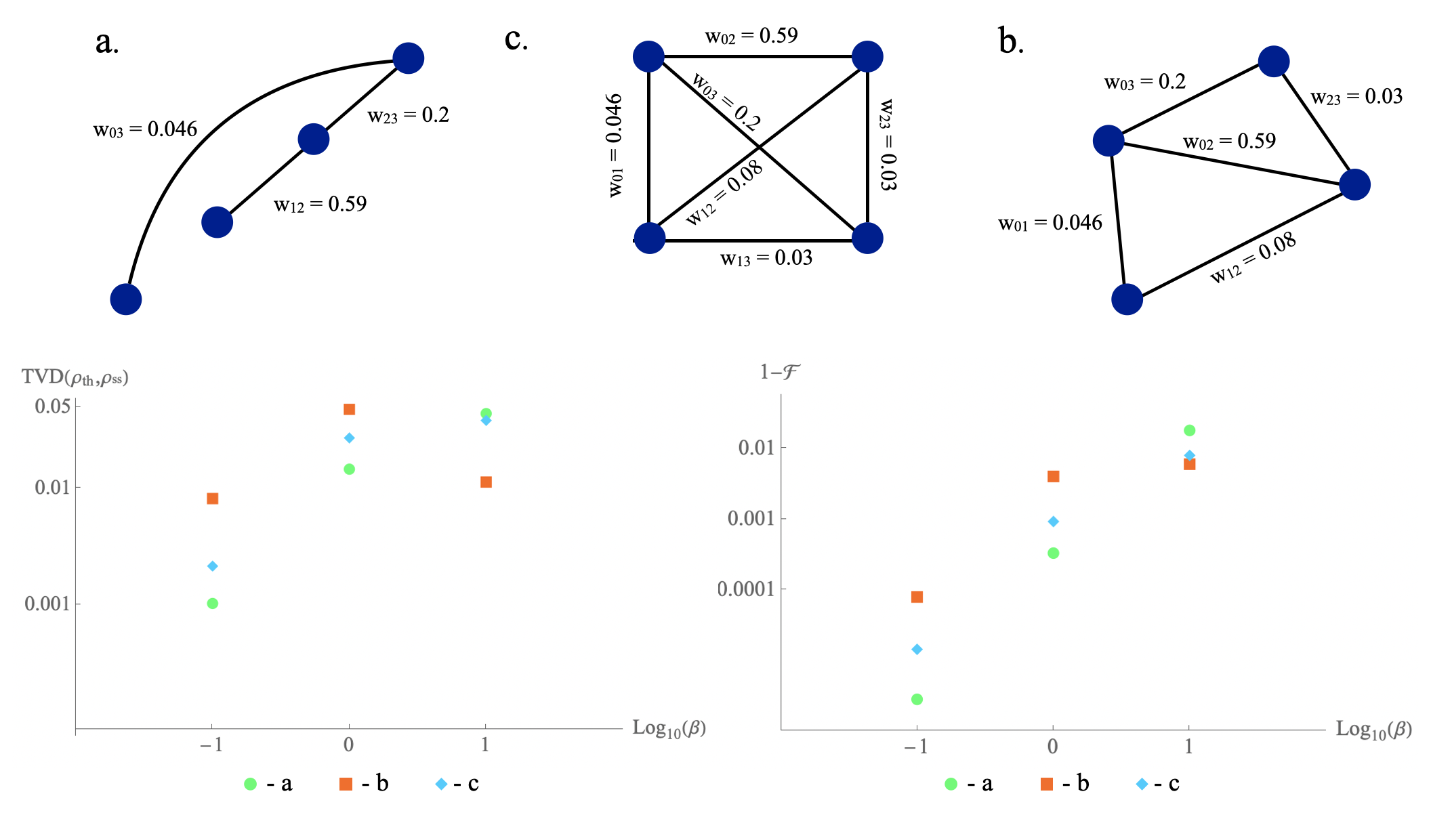}
  \caption{Total variational distance (TVD) and infidelity between the true Gibbs distribution and the distribution generated by the QMCMC algorithm for random instances of graphs with 4 vertices and edge probability a) 0.4, b) 0.6, and c) 1 at inverse temperatures $\beta = 10,1,0.1$. The vertex values are $h_a = \{0.084, 0.026, 0.403, 0.379\}$, $h_b = \{0.403, 0.379, 0.0528, 0.805\}$, $h_c = \{0.379, 0.0528, 0.805, 0.379\}$ and are computed using the same seed as the edges.}
  \label{Fig:RandomGraph}
\end{figure*}

In this section we simulate our QMCMC algorithm on two model applications to determine its performance. The first is a paradigmatic spin model, the transverse-field Ising model (TFIM), and the goal is to prepare the thermal density matrix of the model and reproduce thermal observables. The second example demonstrates our algorithm for the task of sampling from Gibbs distributions determined by probabilistic graphical models over classical random variables.

We determine the performance of our algorithm by numerically calculating the steady state of the dynamical map. To do so, we compute the reduced system dynamics of the system qubits over a complete cycle, or period $T_{\rm cycle}$, of the ancilla sweeping function $f(t)$. The system dynamics is repeated application of this dynamical map defined over the time period $[0,T_{\rm cycle}]$, and therefore the steady state of the system is the zero-eigenvalue eigenvector of the dynamical map over this period. We explicitly compute the dynamical map acting on just the system qubits over this period, defined as:
\begin{align}
	&\rho_s(t+T_{\rm cycle}) \equiv \mathcal{M} \rho(t) \nn \\
	&= \Tr_{b}\left\{\left[\prod_{k=0}^{N_{\rm cycle}}  \mathcal{W}_{t_k} \circ \mathcal{F}_{t_k} \circ \mathcal{R} \right] \left(\rho_s(t)\otimes \ket{0}_b\bra{0}\right) \right\}, \nn
\end{align}
After explicit computation of the linear map $\mathcal{M}$, we compute its spectrum and define the steady state of the dynamics, $\rho_{ss}$ as the  eigenvector associated with eigenvalue one. In all of our numerical studies, this map has a non-degenerate eigenspace at eigenvalue one, and thus a unique steady state. We can also determine the difficulty to achieve thermalization by evaluating the spectral gap of the dynamical map, the inverse of which determines the thermalization time scale \cite{temme_quantum_2011}.

Before we examine the numerical examples, we summarize the expected sources of error in our protocol. Ref. \cite{metcalf2020engineered} presented a detailed analysis of the sources of imperfection in the analog version of our thermalization protocol. There it was shown that the effective spectral density seen by the system can violate detailed balanced conditions due to the finite scale of the ancilla subsystem, and this leads to thermalization errors, especially in systems where the energy gaps in $H_s$ are congested in frequency and when the target temperature is low. We expect poorer thermalization performance for this digital algorithm in that regime also. In addition, the digital algorithm will incur errors due to the Suzuki-Trotter approximation of the analog continuous-time dynamics. 

\textit{Transverse Field Ising Model -- } The one-dimensional TFIM~\cite{de1963collective, stinchcombe1973ising} with open boundary conditions is defined by the Hamiltonian,
\begin{equation}
    H = -J \sum_{i=1}^{N_s-1} \sigma^i_z \sigma^{i+1}_z - h\sum_{i=1}^{N_s} \sigma^i_y,
\end{equation}
which describes spins coupled locally with strength $J$ in the presence of a transverse field $h$. Coupling strength $J$ sets our energy and time scale for the Hamiltonian evolution. Each principal spin is coupled to an ancillary qubit with an interaction strength $g/J = 0.005$ that sweeps the system spectrum with a sinusoidal $f(t) = \sin^2\left(\pi t/T_{\rm cycle}\right)$, with $N_T = 5000$ and $N_{\rm cycle}=500$, implying a period of  $T_{\rm cycle} = T_g N_{\rm cycle} = 10\pi \times 10^4/J$. 

To assess the quality of the thermalization, we compute the infidelity of the steady state of the dynamical map $\mathcal{M}$ defined above with the ideal thermal state at the simulation temperature. The infidelity is defined as $1-\mathcal{F}$, with 
\begin{align}
	\mathcal{F} = \text{Tr}\left(\sqrt{\sqrt{\rho_{th}}\rho_{ss}\sqrt{\rho_{th}}}\right)^2,
\end{align}
where $\rho_{th}$ is the ideal thermal state, which for low temperatures approaches the ground state of the TFIM.

In Fig. \ref{Fig:TFIM_infid} we show the quality of thermalization for TFIM models of varying size and in three regimes: $h/J < 1$, $h/J = 1$, and $h/J > 1$. The target temperature for all simulations is $\beta J=10$. The infidelity generally increases with system size in all parameter regimes. In the regime where $g/J < 1$, the error is significantly greater and increases exponentially with system size. This error arises from the energy structure of the TFIM at low fields. At low fields energy transitions become similar requiring greater resolution of the sweeping function f(t). The steady-state error could be improved in this regime by increasing the resolution of the ancilla sweeping function by increasing $N_{\rm cycle}$.

Finally, for this physically motivated example, we demonstrate the extraction of thermal observables finite-temperature steady-state extracted from the dynamical map spectrum. For the TFIM, a common observable of interest is the magnetization in the transverse field direction, $m_y(t)=\mathrm{tr}(\rho(t) \sigma_y)$, which is used to characterize thermal phase transitions. We evaluate numerically the magnetization for the exact thermal state $\rho_{th}$ and the state $\rho(t)$ obtained from applying algorithm to a random initial state for a number of ancilla sweeps ($T_{\rm cycle}$s). We calculate the magnetization for 1D chain with open boundary conditions where $h/J=1$ for $N_s = 2-5$ principal spins Fig.~\ref{Fig:TFIMMag}.  Interestingly, we see a trend in the error as a function of temperature. Rather than the error steadily increasing with lower temperature, we notice the error is periodic for an even number of qubits as the temperature is increased. Though, we see this pattern emerge for the magnetization, the infidelity of the steady-state increases for low temperatures as expected.

\textit{ Sampling from Gibbs distributions -- }
Sampling from Gibbs distributions of complex networks has a long history in statistical mechanics and machine learning~\cite{Iain2004, FreyGraphs2005, AlbertRev2002}. MCMC techniques are the standard approach to such sampling problems, and here we demonstrate that our QMCMC algorithm can be used for such sampling problems. 

Suppose we have a collection of binary random variables $X_i$, $1\leq i \leq N_S$, whose configurations are dictated by a potential in the form of a quadratic formula in conjunctive normal form, $\eta(X_1, .. N_{N_s})$. We are interested in sampling from the Gibbs distribution defined by $p(X_1, .. N_{N_s}) = \exp(\eta(X_1, ... X_{N_s}))/\mathcal{Z}$, where $\mathcal{Z} = \sum_{X_1,...X_{N_s}}\exp(\eta(X_1, ... X_{N_s}))$ is a partition function. This is equivalent to sampling the Gibbs distribution over a probabilistic graphical model with nodes corresponding to the random variables and each edge corresponding to a clause in the potential formula \cite{koller2009probabilistic}. 

We test the ability of QMCMC to prepare thermal states that enable such Gibbs sampling by encoding the potential in a Hamiltonian of the form
\begin{equation}
    H_g = \sum_{i=1}^{N_s} h_i \sigma_z^i + \sum_{j,k \in \zeta} w_\zeta \sigma_z^j \sigma_z^{k}.
\end{equation}
This Hamiltonian is equivalent to an encoding of $\eta(X_1, ... X_{N_s})$, and the $\ket{0}, \ket{1}$ states of each qubit correspond to the possible values of the random variables. $\zeta$ lists all clauses in $\eta$. In the following, we generate random instances of such problems. To do so, we first generate Erd\"os-R\'enyi random graphs with edge probability $p_e$. These edges in this graph dictate $\zeta$ for the problem. Then for each edge we assign a weight, $w_\zeta$, that is a random variable uniformly distributed in the interval $[0,1]$, and for each variable we assign a random local energy, $h_i$, that is uniformly distributed in the interval $[0,1]$.

After generating such a random instance and its corresponding Hamiltonian, we evaluate the ability of QMCMC to thermalize the many-body system by computing the steady state of the thermalization map $\mathcal{M}$, and comparing the distribution of measurement outcomes in the computational basis to the ideal Gibbs distribution $p(X_1, ... X_{N_s})$. We use the total variation distance (TVD) to quantify the error between these distributions:
\begin{align}
	TVD(p, q) = \frac{1}{2}\sum_{i}\vert p_i - q_i \vert
\end{align}

In Fig. \ref{Fig:RandomGraph} we show the TVDs along with the infidelities achieved by QMCMC with an interaction strength of $g=0.005$ for three random instances with varying $p_e$ for several values of (inverse) temperature. We use the same sinusoidal $f(t)$ used in the TFIM calculations discretized by $N_{\rm cycle} = 100$ and $N_T = 5000$. Distance between the true Gibbs distribution and distribution generated by sampling from the QMCMC steady-state is comparable to the infidelities seen for the TFIM example in the low temperature. Similar to the results for the TFIM, the trend observed in temperature for the physically motivated observable is a feature of the energy structure. Algorithmic accuracy depends on the system energy structure, and we can place no guarantees on the energy structure of a random instance. However, the \textit{trend} seen by infidelity metric as a function of temperature is a good approximation for any given graph.  These infidelity results are in good agreement with the analog protocol, accuracy increases with temperature. 

\section{Conclusion}
Dynamically generating thermal states of quantum systems has historically required modeling a macroscopic environment or obtaining detailed knowledge of the system energy spectra, but in this work we demonstrate a method to represent these systems with linear spatial complexity using time-dependent, ancilla qubits. We analyzed the success of this model using magnetization as an observable in the TFIM and by numerically evaluating the fixed point of evolution for a random graph mapped to the Ising model. When the three criteria (Born-Markov, ergodic dynamics, and detailed balance energy exchange) are met, our algorithm generates a unique steady-state that approximates the thermal state. We aim to use this algorithmic framework to sample other complex distributions and define methods to generate non-equilibrium quantum distributions on quantum computers.

\section*{Acknowledgments}
AFK and ES were supported by the Department of Energy, Office of Basic Energy Sciences, Division of Materials  Sciences and  Engineering under Grant No. DE-SC001946. WAdJ, MM, and KK were supported by the U.S. Department of Energy (DOE) under Contract No. DE-AC02-05CH11231, through the Office of Advanced Scientific Computing Research Accelerated Research for Quantum Computing and Quantum Algorithms Team Programs. MS was supported by the U.S. Department of Energy, Office of Science, Office of Advanced Scientific Computing Research, under the Quantum Computing Application Teams program.

Sandia National Laboratories is a multimission laboratory managed and operated by National Technology \& Engineering Solutions of Sandia, LLC, a wholly owned subsidiary of Honeywell International Inc., for the U.S. Department of Energy's National Nuclear Security Administration under contract DE-NA0003525. This paper describes objective technical results and analysis. Any subjective views or opinions that might be expressed in the paper do not necessarily represent the views of the U.S. Department of Energy or the United States Government. 

\bibliography{main.bib,thermostat.bib}

\end{document}